\documentclass{aes2e}
\usepackage{multirow}
\UseRawInputEncoding
\usepackage{makecell} 
\usepackage{adjustbox}
\usepackage{arydshln}
\usepackage{cite}
\usepackage[colorlinks=true, citecolor=cyan, urlcolor=blue, linkcolor=blue]{hyperref}

\begin{document}
\title{Accelerating Diffusion Transformer-Based Text-to-Speech with Transformer Layer Caching}

\authorgroup{
\author{Siratish Sakpiboonchit$^1$ $^2$}
\email{e1351644@u.nus.edu}
\affil{$^1$National University of Singapore\quad$^2$Botnoi Group}
}

\abstract{\noindent \textbf{Abstract.} This paper presents a method to accelerate the inference process of diffusion transformer (DiT)-based text-to-speech (TTS) models by applying a selective caching mechanism to transformer layers. Specifically, I integrate SmoothCache into the F5-TTS architecture, focusing on caching outputs of self-attention and feed-forward network layers to reduce redundant computations during the denoising process. A calibration phase is introduced to analyze L1 relative errors between timesteps, guiding the selection of cache schedules that minimize quality degradation. To address the problem of inter-layer dependency, a unified caching schedule is adopted, applying the cache pattern derived from self-attention layers to both layer types. Experiments on LibriSpeech-PC and Seed-TTS datasets evaluate various cache thresholds and denoising step configurations. Results show that caching at higher denoising steps reduces inference time without compromising output quality, whereas caching at lower steps can negatively impact synthesis quality similarly to reducing the total number of denoising steps. Objective and subjective metrics confirm the effectiveness of SmoothCache in maintaining performance while improving computational efficiency. Comparisons between cached inference and reduced-step inference further highlight the benefits of selective caching, especially under high-step configurations. This work demonstrates that transformer layer caching is a practical solution for optimizing diffusion transformer-based TTS models without requiring architectural changes or retraining. Example inference results can be heard at \url{https://siratish.github.io/F5-TTS_SmoothCache/}.}
\maketitle
\section{Introduction}

\noindent Text-to-Speech (TTS) systems have recently achieved remarkable progress \cite{wang2023viola-49e,tan2024naturalspeech-bf0}, particularly with the rise of diffusion-based models capable of generating high-fidelity, natural-sounding speech conditioned on just a few seconds of prompt audio \cite{shen2023naturalspeech-1c9,le2023voicebox-195,ju2024naturalspeech-34b}.

TTS models can be broadly categorized into autoregressive (AR) and non-autoregressive (NAR) frameworks. AR models, such as Tacotron 2 \cite{shen2018natural-5b8}, generate speech sequentially, conditioning each frame on previous outputs. While this allows strong modeling of temporal dependencies, it results in slow inference and error accumulation during synthesis \cite{han2024vall-e-39a,peng2024voicecraft-ded,song2025ella-v-f8f}. In contrast, NAR models, such as FastPitch \cite{lancucki2021fastpitch-701}, produce all frames in parallel, enabling faster synthesis without sequential dependencies. However, this parallelism introduces challenges in modeling long-range dependencies and necessitates explicit alignment mechanisms to ensure accurate mapping between input text and output speech \cite{ju2024naturalspeech-34b,le2023voicebox-195,mehta2024matcha-tts-8ef}.

Diffusion-based TTS models have emerged as an alternative that combines NAR efficiency with high synthesis quality \cite{shen2023naturalspeech-1c9,ju2024naturalspeech-34b}. These models iteratively denoise latent audio representations, gradually converting noise into natural speech. While they avoid the alignment complexity of traditional NAR models, they remain computationally intensive due to repeated denoising steps across many timesteps.

Early examples such as E3-TTS \cite{gao2023e3-277} attempted to eliminate explicit phoneme-level alignment using cross-attention over character inputs but achieved limited audio quality. DiTTo-TTS used Diffusion Transformer (DiT) \cite{peebles2023scalable-137} with cross-attention conditioned on encoded text and improved synthesis fidelity by using pretrained language model to fine-tune the neural audio codec. E2-TTS \cite{eskimez2024e2-f63} (based on Voicebox \cite{le2023voicebox-195}) further simplified the pipeline by directly using padded character sequences as input and removing phoneme and duration predictors, although alignment robustness remained a concern.

\begin{figure}[h]
    \centering
    \includegraphics[width=0.8\linewidth]{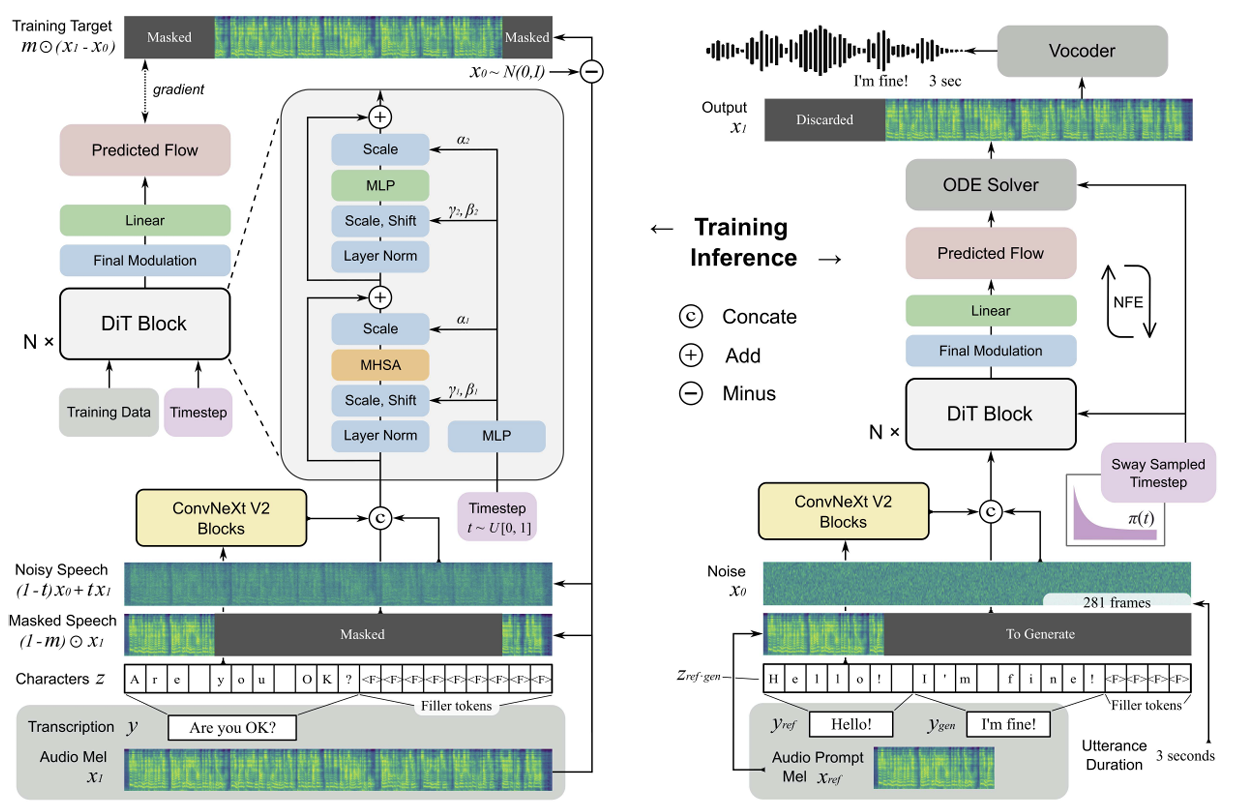}
    \caption{Overview of F5-TTS training (left) and inference (right). The model is trained via text-guided speech infilling with Flow Matching loss. Input text is converted to padded character sequences and refined via ConvNeXt V2 before concatenation with speech input. Inference employs Sway Sampling to generate speech from noise.}
    \label{fig:F5-TTS}
\end{figure}

Building upon these, F5-TTS \cite{chen2024f5-tts-7fa} offers a streamlined yet effective approach. It discards explicit alignment modules and phoneme modeling by directly embedding padded character sequences via ConvNeXt V2 \cite{woo2023convnext-277} and conditioning a DiT model with Flow Matching \cite{lipman2022flow-819}. Its “Sway Sampling” inference strategy dynamically adjusts the sampling process for faster generation. F5-TTS achieves a real-time factor (RTF) of approximately 0.15 while maintaining high naturalness, speaker similarity, and intelligibility. An overview of the F5-TTS architecture is illustrated in Fig.\ref{fig:F5-TTS}.

Despite these advancements, F5-TTS, like other DiT-based models, remains constrained by the computational burden of iterative denoising. Prior work has explored various optimization strategies, including advanced solvers \cite{liu2022flow-6c6,lu2025dpm-solver-260}, knowledge distillation \cite{salimans2022progressive-800}, pruning \cite{fang2023structural-42c}, and quantization \cite{he2023ptqd-6b9,shang2023post-training-66a}.

\begin{figure}[h]
    \centering
    \includegraphics[width=0.8\linewidth]{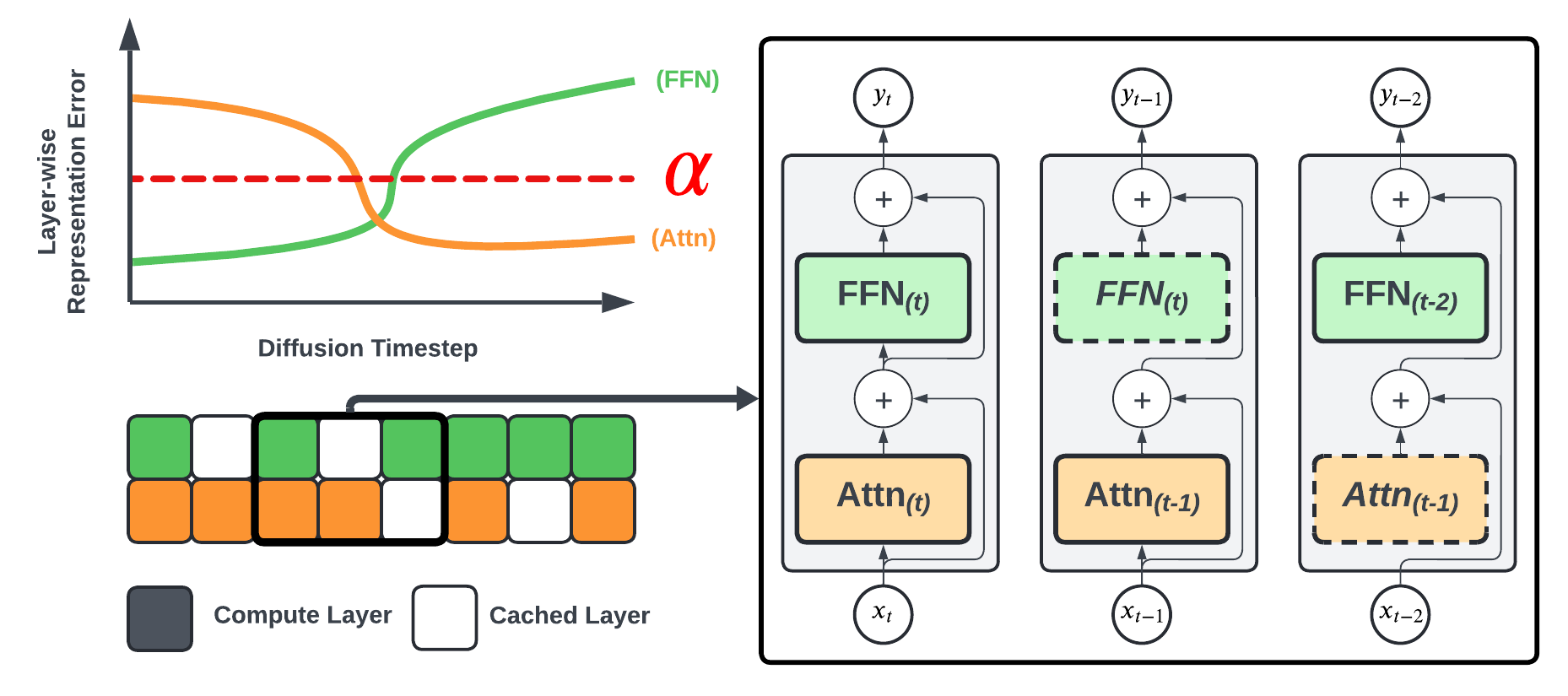}
    \caption{Illustration of SmoothCache. During a calibration pass, layer-wise representation errors are analyzed. If the error falls below a threshold $\alpha$, the layer output is cached for reuse in future timesteps. The left side shows the representation error trend; the right side depicts SmoothCache applied to a DiT model using residual connections to reintroduce cached outputs.}
    \label{fig:SmoothCache}
\end{figure}

A particularly promising zero-training, inference-only method is SmoothCache \cite{liu2024smoothcache-ba8}, an optimization technique designed to accelerate inference of DiT models. It exploits the observation that hidden representations in transformer layers exhibit redundancy across consecutive denoising timesteps. A short calibration pass measures cosine similarities to determine which layers and timesteps are safe to cache. These cached outputs are reused during inference, reducing computation without altering model parameters (see Fig.\ref{fig:SmoothCache} for schematic). While SmoothCache has demonstrated effectiveness in accelerating diffusion models for images, video, and audio generation, it has not yet been applied to TTS tasks.

In this project, I investigate integrating SmoothCache into the F5-TTS pipeline without retraining the model. I evaluate caching strategies and assess their impact on inference speed and synthesis quality. Experimental results demonstrate that SmoothCache enables substantial inference acceleration with minimal degradation in output quality.

\section{Data}

\noindent This project uses the same evaluation datasets as the F5-TTS paper: 1) LibriSpeech-PC \textit{test-clean} \cite{meister2023librispeech-pc-d85}, and 2) Seed-TTS \textit{test-en} \cite{anastassiou2024seed-tts-5b7}.

\subsection{LibriSpeech-PC \textit{test-clean}}

\noindent A filtered subset of LibriSpeech test-clean ($\sim$1000h, 16 kHz read English speech) focused on 4-10 second utterances. This subset contains 1,127 audio-text pairs ($\sim$2h total), drawn from 39 speakers. Audio is stored in FLAC format, named as \texttt{speakerID-chapterID-utteranceID.flac}. Corresponding transcripts are stored in per-chapter \texttt{.txt} files, where each line follows the format:
\begin{verbatim}
speakerID-chapterID-utteranceID TRANSCRIPTION_TEXT
\end{verbatim}
The F5-TTS repository\footnote{\url{https://github.com/SWivid/F5-TTS}} also provides a \texttt{.lst} metadata file, in which each row consists of six tab-separated fields:
\begin{verbatim}
ref_utt ref_dur ref_txt gen_utt gen_dur gen_txt
\end{verbatim}
Each field is defined as follows:
\begin{itemize}
  \item \textbf{ref\_utt}: Identifier of the reference utterance (e.g. \texttt{1234-5678-0001})
  \item \textbf{ref\_dur}: Duration of the reference audio in seconds 
  \item \textbf{ref\_txt}: Text transcript of the reference utterance 
  \item \textbf{gen\_utt}: Identifier assigned to the generated utterance (e.g. \texttt{1234-5678-0002})
  \item \textbf{gen\_dur}: Duration of the generated audio in seconds 
  \item \textbf{gen\_txt}: Text transcript used to generate the synthesized speech
\end{itemize}
Reference and generated utterances are paired via a cross-sentence sampling strategy, enabling objective quality evaluations.

\subsection{Seed-TTS \textit{test-en}}

\noindent This test set is derived from Common Voice (English) \cite{ardila2019common-b5d}, consisting of 1,088 utterances. Audio is explicitly split into reference and generated files. Filenames follow:
\begin{itemize}
  \item \texttt{<ref\_utt>.wav} — Reference Audio
  \item \texttt{<ref\_utt>-<gen\_utt>.wav} — Generated Audio
\end{itemize}
Metadata is provided in a \texttt{.lst} file, with each row containing:
\begin{verbatim}
utt prompt_text prompt_wav gt_text gt_wav
\end{verbatim}
Fields are defined as follows:
\begin{itemize}
  \item \textbf{utt}: Unique identifier for the utterance pair (e.g. \texttt{common\_voice\_en\_00001-common\_voice\_en\_00002})
  \item \textbf{prompt\_text}: Text transcript of the reference prompt 
  \item \textbf{prompt\_wav}: File path to the reference audio (\texttt{.wav})
  \item \textbf{gt\_text}: Text transcript intended for generation
  \item \textbf{gt\_wav}: File path to the generated audio (\texttt{.wav}), if available
\end{itemize}
This metadata structure supports text-content alignment and separate access to reference vs. synthetic audio for evaluation.

\section{Methods and Algorithms}

\noindent In this project I integrate SmoothCache into the F5-TTS inference pipeline in two stages: a Calibration Phase to discover a robust cache schedule, and an Inference Phase to apply that schedule and accelerate synthesis.

\subsection{Calibration Phase}

\begin{figure}[h]
  \centering
  \includegraphics[width=0.8\linewidth]{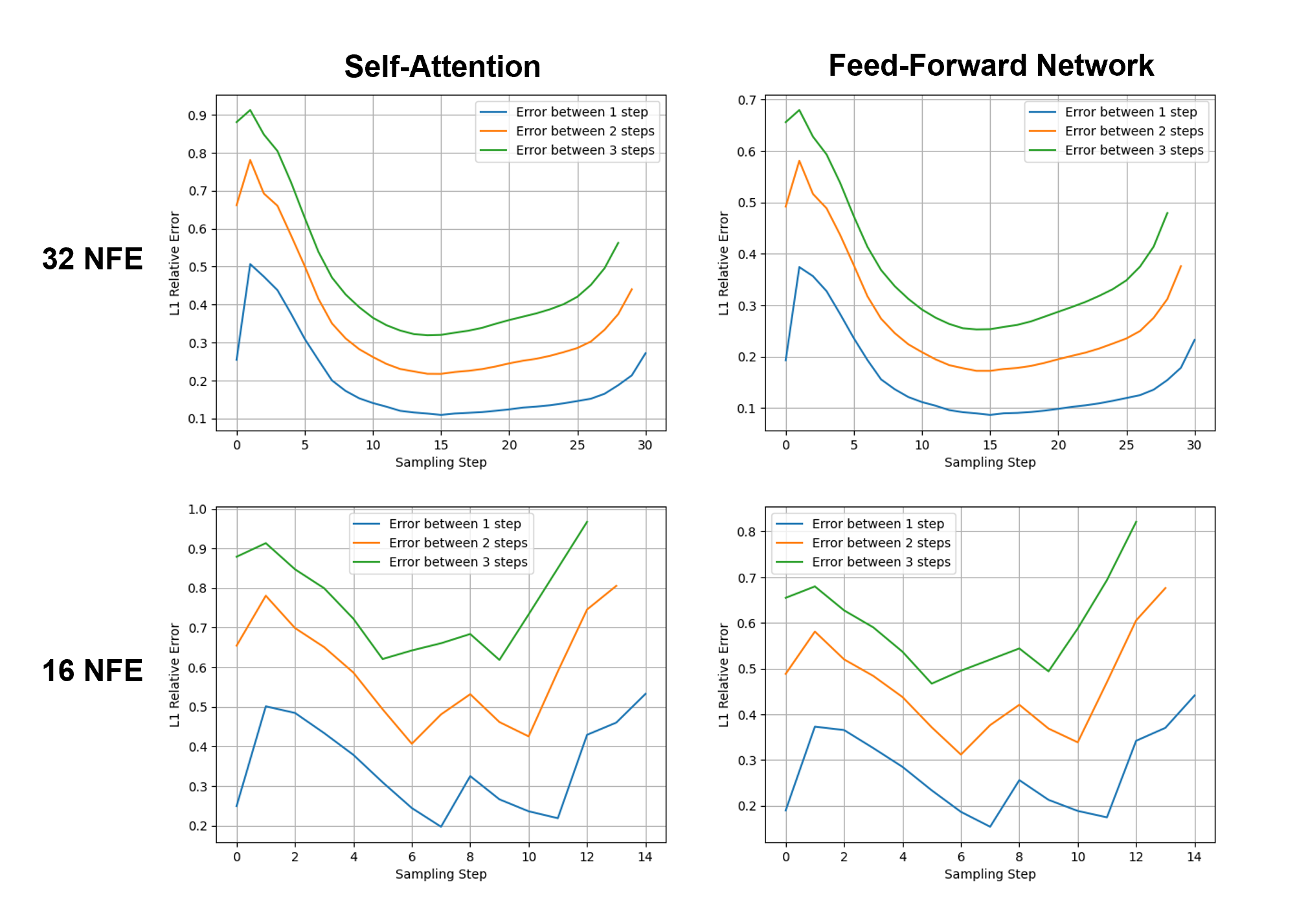}
  \caption{L1 relative error between consecutive timesteps for self-attention and FFN layers, averaged over 10 samples. Curves are scaled to the same y-axis range.}
  \label{fig:curve}
\end{figure}

\noindent In the Calibration Phase, I focus on the two most computationally intensive sub-layers of the Diffusion Transformer backbone: the self-attention (Attn) block and the feed-forward network (FFN) block, each followed by a residual connection. I process a few audio samples through the standard F5-TTS denoising loop and record the L1 relative error between consecutive timesteps for each layer. The resulting error curves, shown in Fig.\ref{fig:curve}, highlight three trends: large errors at the beginning and end of the diffusion trajectory, consistently higher errors in Attn than in FFN, and an elevated baseline error when the number of steps is reduced. Using these insights, I select the default caching threshold \(\alpha\) and generate an initial cache schedule for Attn and FFN.

\begin{figure}[h]
  \centering
  \includegraphics[width=0.8\linewidth]{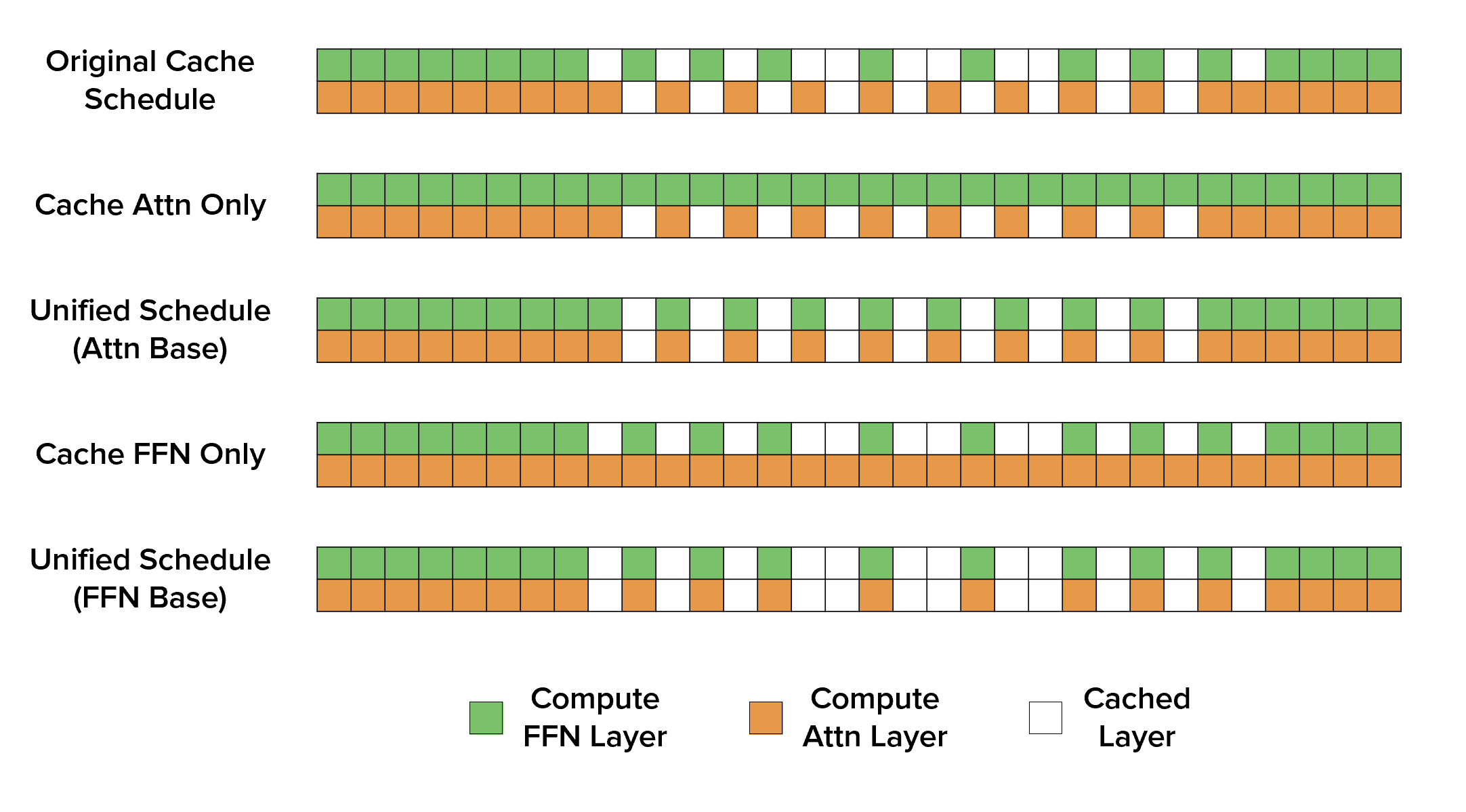}
  \caption{Different caching strategies explored during calibration. The original cache schedule was derived using a threshold of \(\alpha=0.15\). Example inference results for each approach are available on the demo page.\setcounter{footnote}{1}\protect\footnotemark}
  \label{fig:calibrate}
\end{figure}

However, the original cache schedules, derived by evaluating each layer in isolation, suffer from a key limitation noted in the SmoothCache paper: they do not account for interactions between layers when both are cached simultaneously \cite{liu2024smoothcache-ba8}. In practice, applying these per-layer schedules caused noticeable audio artifacts, as caching one layer without recomputing the other disrupted the residual connections and degraded synthesis quality. To diagnose this, I conducted listening tests on a single audio example using five different caching strategies as shown in Fig.\ref{fig:calibrate}: the original independent schedules, caching only Attn, caching only FFN, and two unified schedules that applies the pattern of one layer to both layers (Attn base and FFN base). These tests showed that “Attn-Only” introduced fewer artifacts than “FFN-Only”, and that the unified (Attn base) schedule preserved Attn-Only audio fidelity while maximizing cache utilization. Consequently, I adopted this unified schedule for all subsequent experiments.

\subsection{Inference Phase}
\noindent In the Inference Phase, given a reference audio prompt (mel-spectrogram $x_{\mathrm{ref}}$) and its transcription $y_{\mathrm{ref}}$, along with a target text prompt $y_{\mathrm{gen}}$, I inject the unified (Attn base) cache schedule into the F5-TTS denoising loop. At each diffusion step $t$, the model first denoises via an ODE solver with Sway Sampling. Before each residual connection in the Diffusion Transformer, I check whether $t$ is marked for caching: if so, the corresponding layer output (Attn or FFN) is retrieved from the cache; otherwise, it is computed normally and the cache is updated. After completing $T$ steps, the final mel-spectrogram $x_{\mathrm{gen}}$ is converted into a waveform using the vocoder.
\footnotetext{\url{https://siratish.github.io/F5-TTS_SmoothCache/}}
\subsection{Experimental Setup}

\noindent\textbf{F5-TTS Setup}\quad
I use the publicly available F5-TTS v1 base model\footnote{\url{https://huggingface.co/SWivid/F5-TTS}} with exponential moving averaged (EMA) \cite{karras2024analyzing-381} weights. Inference is performed with the Euler ODE solver, a CFG \cite{ho2022classifier-free-d8a} strength of 2, and a Sway Sampling coefficient of \(-1\). The final log mel spectrograms are converted to waveform via the pretrained vocoder Vocos \cite{siuzdak2023vocos-faf}. All hyperparameters and solvers match those in the original F5-TTS paper to ensure a controlled baseline.

\begin{figure}[h]
    \centering
    \includegraphics[width=0.8\linewidth]{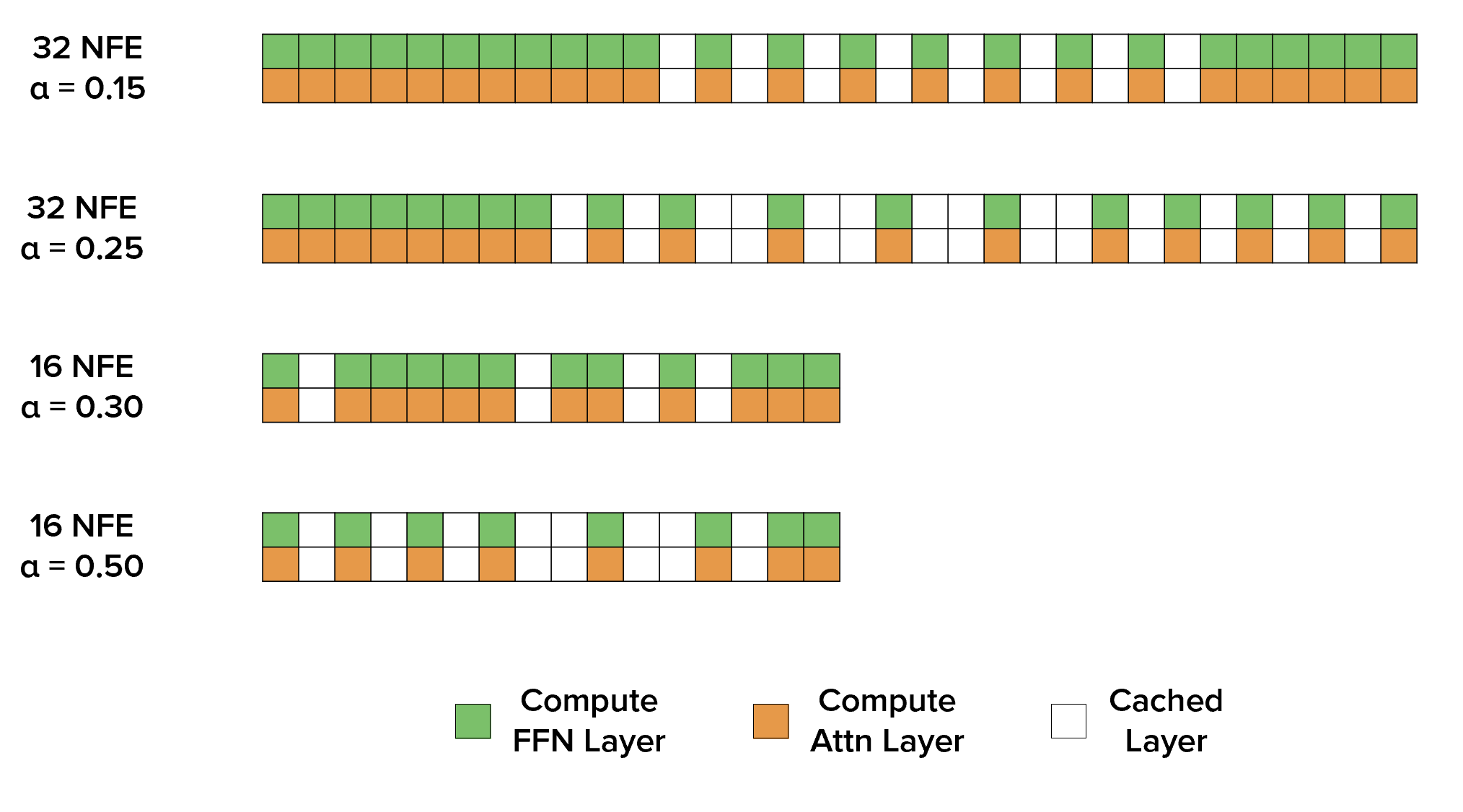}
    \caption{Cache schedules with different NFE steps and caching thresholds (\(\alpha\)) used for experiments. All schedules follow the Unified (Attn base) approach.}
    \label{fig:cache}
\end{figure}

\noindent\textbf{SmoothCache Setup}\quad
For calibration, I randomly sample ten utterances from LibriSpeech-PC \textit{test-clean}. I constrain the cache schedule so that no more than three consecutive timesteps are cached per layer, preventing excessive approximation error. Cache schedules for all experiments are as shown in Fig.\ref{fig:cache}. For 32 NFE, I compare the baseline to two thresholds: $\alpha=0.15$, which results in caching approximately one-quarter of all steps, and $\alpha=0.25$, which caches around half of the steps. For 16 NFE, I experiment with $\alpha=0.30$ and $\alpha=0.50$ to obtain cache schedules covering comparable proportions of timesteps. Those schedules are applied across the entire LibriSpeech-PC \emph{test-clean} and Seed-TTS \emph{test-en} datasets for evaluation.

\noindent\textbf{Evaluation Metrics}\quad
Objective evaluation includes Word Error Rate (WER), measured by transcribing generated speech with Whisper-large-v3 \cite{radford2022robust-174}, and Speaker Similarity (SIM-o), computed as the cosine similarity between WavLM-large \cite{chen2022large-scale-55e} speaker embeddings of synthesized and ground-truth audio. For subjective evaluation, I record predicted mean opinion scores using two open-source MOS predictors: UTMOS strong \cite{saeki2022utmos-4f1} and NISQA v2.0 (mos only) \cite{mittag2021nisqa-2a9}. I also conduct a user study with 37 participants, each judging 30 randomly selected utterance pairs in A/B trials on speaker similarity and overall quality. More details of the user study on Sec.\ref{4.1}.

All evaluations follow the cross-sentence protocol \cite{chen2025neural-533,le2023voicebox-195} on LibriSpeech-PC \textit{test-clean} and Seed-TTS \textit{test-en}. I report the average scores across three random seed generation results for each experiment to account for stochastic variability.

\begin{Table*}
\centering
\tbl{\label{tab:1} Comparison results of F5-TTS on LibriSpeech-PC \textit{test-clean} and Seed-TTS \textit{test-en}. The Real-Time Factor (RTF) is computed with the inference time of 10s speech on The NVIDIA T4 GPU.}{%
\begin{tabular}{l|cccc|cccc|c}
\Xhline{1.1pt} 
\multicolumn{1}{c}{}& \multicolumn{4}{c}{\textbf{LibriSpeech-PC \textit{test-clean}}} & \multicolumn{4}{c}{\textbf{Seed-TTS \textit{test-en}}} &  \\
\multicolumn{1}{l}{\textbf{Schedule}} & \textbf{WER(\%)$\downarrow$} & \textbf{SIM-o$\uparrow$} & \textbf{UTMOS$\uparrow$} & \multicolumn{1}{c}{\textbf{NISQA$\uparrow$}} & \textbf{WER(\%)$\downarrow$} & \textbf{SIM-o$\uparrow$} & \textbf{UTMOS$\uparrow$} & \multicolumn{1}{c}{\textbf{NISQA$\uparrow$}} & \textbf{RTF}$\downarrow$ \\
\hline
32 NFE (No Cache)     & 2.03 & 0.67 & 3.86 & 4.23 & 1.59 & 0.68 & 3.67 & 3.99 & 0.46 \\
32 NFE ($\alpha=0.15$)& 2.05 & 0.68 & 3.86 & 4.21 & 1.58 & 0.68 & 3.69 & 4.01 & 0.32 \\
32 NFE ($\alpha=0.25$)& 2.06 & 0.68 & 3.87 & 4.19 & 1.52 & 0.68 & 3.71 & 4.01 & 0.26 \\
\hdashline
16 NFE (No Cache)     & 2.05 & 0.68 & 3.90 & 4.14 & 1.54 & 0.69 & 3.74 & 3.97 & 0.23 \\
16 NFE ($\alpha=0.30$)& 2.02 & 0.68 & 3.87 & 4.09 & 1.55 & 0.69 & 3.74 & 3.96 & 0.18 \\
16 NFE ($\alpha=0.50$)& 2.17 & 0.67 & 3.79 & 3.97 & 1.56 & 0.68 & 3.72 & 3.91 & 0.12 \\
\Xhline{1.1pt} 
\end{tabular}}%
\end{Table*}

\section{Discussion}

\noindent Table~\ref{tab:1} illustrates that applying SmoothCache to F5-TTS yields substantial inference speedups with minimal quality loss. For high NFE settings (e.g., 32 steps), caching approximately one quarter of the layers (\(\alpha=0.15\)) reduces the RTF from 0.46 to 0.32, and caching half of the layers (\(\alpha=0.25\)) further lowers RTF to 0.26, nearly a 2× speedup. Even at lower NFE (16 steps), caching accelerates inference (RTF from 0.23→0.18 for \(\alpha=0.30\)), though overly aggressive caching (\(\alpha=0.50\)) can degrade performance by eliminating too many compute steps.

A closer examination of the two evaluation datasets reveals subtle yet informative differences. On LibriSpeech-PC \emph{test-clean}, which comprises read audiobook speech with consistent pacing and clarity, all caching configurations maintain nearly identical WER and speaker similarity scores. As the proportion of cached steps increases, subjective scores decline slightly but predictably, remaining within a narrow range. In contrast, Seed-TTS \emph{test-en} encompasses a broader variety of speaking styles, speeds, and recording conditions, resulting in some inconsistency in perceptual metrics. For instance, NISQA scores on Seed-TTS improve marginally (3.99→4.01) at \(\alpha=0.15\) and 0.25, likely because the neural codec’s resilience to diverse inputs masks minor caching artifacts; however, larger thresholds (e.g., \(\alpha=0.50\)) produce the expected slight quality dips (3.97→3.91). These observations underscore that while SmoothCache delivers consistent speed-quality trade-offs on uniform datasets, it is less straightforward for more varied corpora.

\begin{Table*}
\centering
\tbl{\label{tab:2} Comparison results of F5-TTS on Seed-TTS \textit{test-en} under two strategies: (1) applying SmoothCache while maintaining the original number of NFE steps, and (2) reducing the number of NFE without caching, matched to the same number of compute steps. Includes results from a user study evaluating both settings.}{%
\begin{tabular}{l|ccccc|c}
\Xhline{1.1pt} 
\multicolumn{1}{l}{\textbf{Schedule}} & \textbf{WER(\%)$\downarrow$} & \textbf{SIM-o$\uparrow$} & \textbf{UTMOS$\uparrow$} & \textbf{NISQA$\uparrow$} & \multicolumn{1}{c}{\textbf{RTF}$\downarrow$} & \textbf{User Study} \\
\hline
32 NFE ($\alpha=0.15$)& 1.58 & 0.68 & 3.69 & 4.01 & 0.32 & 53\% \\
24 NFE (No Cache)     & 1.54 & 0.68 & 3.68 & 3.99 & 0.31 & 47\% \\
\hdashline
16 NFE ($\alpha=0.30$)& 1.55 & 0.69 & 3.74 & 3.96 & 0.18 & 49\% \\
12 NFE (No Cache)     & 1.59 & 0.68 & 3.72 & 3.96 & 0.16 & 51\% \\
\Xhline{1.1pt} 
\end{tabular}}%
\end{Table*}

\subsection{Ablation of Caching Steps} \label{4.1}
\noindent Having established that SmoothCache can effectively optimize F5-TTS inference time without significant degradation in quality, I further investigate whether caching steps genuinely preserve synthesis performance compared to simply skipping those steps entirely. Specifically, according to Fig.\ref{fig:cache}, since 32 NFE with caching threshold $\alpha=0.15$ results in 24 compute steps (due to 8 cached steps), it is natural to compare this setup with a 24 NFE baseline without caching. Similarly, 16 NFE with $\alpha=0.30$ yields 12 compute steps, motivating its comparison against a 12 NFE baseline without caching.

Table~\ref{tab:2} summarizes the evaluation results from comparing these paired configurations. Across both pairs, all objective and subjective metrics appear closely matched. However, the user study results provide additional subjective insight: in the 32 NFE comparison, the cached version slightly outperforms the 24 NFE no-cache variant (53\% vs. 47\% preference), suggesting a marginal perceptual benefit from using cached steps rather than eliminating them outright. In contrast, the 16 NFE pair shows almost equal performance, indicating little distinction between caching and step reduction in this lower NFE setting.

These findings suggest that applying caching at higher NFE (longer generation trajectories) is more advantageous than reducing NFE steps directly. The reason may be that higher NFE schedules inherently possess more redundant or less impactful steps suitable for caching without affecting quality. In contrast, lower NFE schedules, which are already condensed, have fewer unimportant steps available for caching, meaning that caching at this level can harm synthesis quality similarly to reducing NFE steps. This analysis reinforces that SmoothCache is most beneficial when applied to higher NFE schedules, where it offers runtime acceleration without compromising output quality.

\subsection{Limitations and Future Work}
\noindent The primary limitation of this work lies in the restricted scope of experiments due to computational cost and time constraints. While the results demonstrate that SmoothCache effectively reduces inference time without significantly degrading output quality, there remain several unexplored avenues that could lead to further improvements.

In the calibration process, although applying the cache schedule derived from the Attn layer to both Attn and FFN layers successfully mitigated dependency issues, this approach is admittedly heuristic. A more principled strategy could involve integrating L1 relative error curves from both layers when constructing cache schedules, potentially weighting the contribution from each layer or treating them equally. Additionally, using alternative datasets or different sampling strategies for calibration could reveal whether the observed error distributions are model- or dataset-specific.

Regarding the inference process, the current experiments are limited to varying the number of denoising steps (NFE) and caching thresholds. Future work should investigate how SmoothCache interacts with other inference hyperparameters, such as CFG strength, Sway Sampling coefficient, and different ODE solvers (e.g., midpoint or Heun-3). Similarly, exploring its performance with different vocoders, such as BigVGAN \cite{lee2022bigvgan-259}, could provide broader insights into its generalizability across synthesis pipelines.

In terms of evaluation, the current subjective assessment is limited to direct A/B preference tests comparing output pairs. Expanding user studies to include tests focused on specific perceptual attributes could yield richer insights. For instance, conducting CMOS tests (comparing each generated sample against a reference and rating from -3 to +3) would provide quantitative data on overall audio quality relative to natural speech. Similarly, SMOS tests (speaker similarity ratings from 1 to 5) could offer a more focused measure of speaker identity preservation.

\section{Conclusion}

\noindent This study investigated the application of SmoothCache to accelerate F5-TTS inference by selectively caching transformer layer outputs without modifying or retraining the base model. Through a calibration process based on L1 relative error analysis and a unified caching strategy to address inter-layer dependencies, SmoothCache achieved significant reductions in inference time while maintaining synthesis quality. Experiments on LibriSpeech-PC and Seed-TTS demonstrated that caching at higher NFE steps preserves quality more effectively than reducing denoising steps, whereas caching at lower NFE levels degrades performance comparably to step reduction. The results suggest that calibrated caching offers a practical approach to optimizing diffusion-based TTS models. Future work could enhance calibration strategies, explore more varied inference conditions, and refine evaluation methodologies to extend the applicability and effectiveness of SmoothCache.
\newpage
\bibliography{references}
\bibliographystyle{vancouver}

\end{document}